\documentclass{article}
\usepackage[utf8]{inputenc}
\usepackage{amsmath}
\usepackage{amssymb}
\usepackage{booktabs}
\usepackage{caption}
\usepackage{graphicx}
\usepackage{geometry}
\usepackage[numbers,super,sort&compress]{natbib}
\usepackage{xurl}
\title{A Dataset of Equilibrium State Configurations of Adsorption in Zeolites}
\author{%
Marko Petkovic\thanks{Department of Applied Physics and Science Education, Eindhoven University of Technology, Eindhoven, the Netherlands.}
\and Rachna Ramesh\footnotemark[1]
\and Vlado Menkovski\thanks{Department of Mathematics and Computer Science, Eindhoven University of Technology, Eindhoven, the Netherlands.}
\and Sofia Calero\footnotemark[1]\thanks{Corresponding author: \texttt{s.calero@tue.nl}.}%
}
\date{}

\begin{document}

\maketitle

\begin{abstract}
Zeolites are crystalline nanoporous materials widely used in adsorption, separation, and catalytic processes.  Molecular simulations are commonly used to predict adsorption properties, but most high-throughput adsorption datasets report only ensemble-averaged quantities such as loadings or isotherms, rather than the molecular configurations from which these averages are obtained. Here, we present AdsZeo, a coordinate-resolved dataset of equilibrium methane adsorption configurations in aluminium-substituted, sodium-containing zeolite frameworks. The processed release contains 4,775 framework realisations derived from 191 zeolite topologies. Each framework realisation was simulated at 13 methane pressures between 0.1 and 100 bar at 298 K using grand canonical Monte Carlo simulations, giving 62,075 production simulations in total. In addition to scalar adsorption records, the dataset stores production-frame methane pseudo-atom coordinates, mobile Na$^+$ cation coordinates, framework atomic coordinates, per-frame loading and energy statistics, and simulation metadata in a processed DuckDB database. The release contains 12,415,000 saved production-frame records and 1,245,376,215 saved particle-coordinate records. AdsZeo provides coordinate-resolved adsorption data across variations in framework topology, aluminium content and distribution, sodium cation arrangement, pressure, and methane loading, enabling reuse for adsorption analysis, spatial statistics, density estimation, and machine-learning models for molecular configuration generation in charged zeolite pores.

\end{abstract}

\section{Background \& Summary}

Nanoporous crystalline materials, such as zeolites and Metal-Organic Frameworks (MOFs), are widely studied for adsorption, separation, gas storage, and catalytic processes \cite{perez2022zeolites,lee2009metal,li2019porous}. Predicting adsorption in these materials requires sampling molecular configurations under specified thermodynamic conditions, because adsorption properties are ensemble averages over many possible arrangements of guest molecules, framework atoms, and extra-framework species.  Configuration-level adsorption data are also needed for machine learning models that aim to emulate equilibrium states, generate adsorbate configurations, or accelerate molecular simulations, rather than only predict scalar adsorption properties. Grand canonical Monte Carlo (GCMC) simulations are therefore a standard tool for computing adsorption isotherms, Henry coefficients, heats of adsorption, and selectivities in porous materials \cite{duren2009using,dubbeldam2016raspa}. In GCMC, molecular configurations are sampled in the grand-canonical ensemble through Monte Carlo moves such as insertions, deletions, translations, rotations, and reinsertion moves, allowing finite-temperature adsorption equilibria to be estimated at fixed chemical potential, volume, and temperature \cite{dubbeldam2013inner}. Compared with electronic-structure methods such as density functional theory, which are typically too expensive for extensive finite-temperature sampling over many pressures, compositions, and frameworks, GCMC provides a practical route to high-throughput adsorption prediction. However, GCMC can still be computationally demanding, especially for systems with low insertion or move acceptance rates, such as dense adsorbate phases, strongly interacting molecules including water, heterogeneous adsorption sites, or flexible frameworks \cite{siderius2024flat,yu2021incorporating}.

Machine learning offers a route to reduce this sampling cost. Deep generative models are well suited to learning complex, high-dimensional probability distributions and have demonstrated substantial success in applications including text generation \cite{brown2020language}, image synthesis \cite{rombach2022high}, and protein design \cite{watson2023novo}. In molecular simulation, Boltzmann generators \cite{noe2019boltzmann} were developed specifically to generate samples from equilibrium distributions and have been applied to molecular conformations \cite{jing2022torsional}, molecular dynamics trajectories \cite{klein2023equivariant}, and peptide systems \cite{klein2024transferable}. These studies demonstrate the potential of generative models for equilibrium sampling, but they have mostly focused on individual molecular systems or relatively small biomolecular examples. Extending such approaches to adsorption in nanoporous materials introduces additional sources of complexity, because the sampled configuration space depends on host frameworks containing hundreds to thousands of atoms, adsorbate loading, pressure, and, in charged frameworks, mobile extra-framework species. For adsorption in porous materials, generative models could learn equilibrium distributions of adsorbates and extra-framework cations, propose more efficient Monte Carlo moves, initialise simulations closer to equilibrium, or approximate adsorbate density fields directly. However, these models require training data at the level of molecular configurations rather than only scalar adsorption properties.

Most existing high-throughput adsorption datasets are not designed to store the molecular configurations generated during adsorption simulations. Instead, they usually report averaged adsorption properties, such as loadings, isotherms, selectivities, Henry coefficients, or heats of adsorption \cite{bobbitt2023mofx,boyd2019data,altintas2018database,oliveira2023crafted}. These properties are essential for comparing materials and training property-prediction models, but they do not retain the individual adsorbate configurations from which the averages were computed. Recent datasets such as Open DAC do include atomistic MOF--adsorbate structures \cite{sriram2024open,sriram2025open}, but these are based on selected adsorbate placements and subsequent relaxation, often involving one or a small number of adsorbate molecules. They are therefore different from datasets that store pressure-dependent equilibrium configurations sampled across adsorption isotherms. As a result, there is a shortage of curated adsorption datasets that expose equilibrium configurations for training and benchmarking models that generate molecular adsorption configurations.

\begin{figure}
    \centering
    \includegraphics[width=0.95\linewidth]{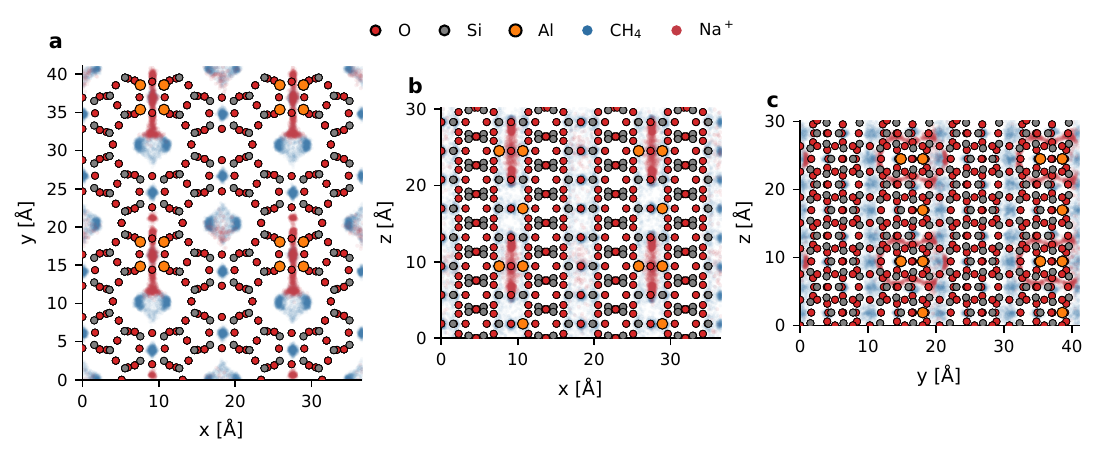}
    \caption{Spatial distribution of methane molecules and sodium cations within a mordenite zeolite framework.}
    \label{fig:mor_dist}
\end{figure}

\begin{figure}[htbp]
\includegraphics[width=\textwidth]{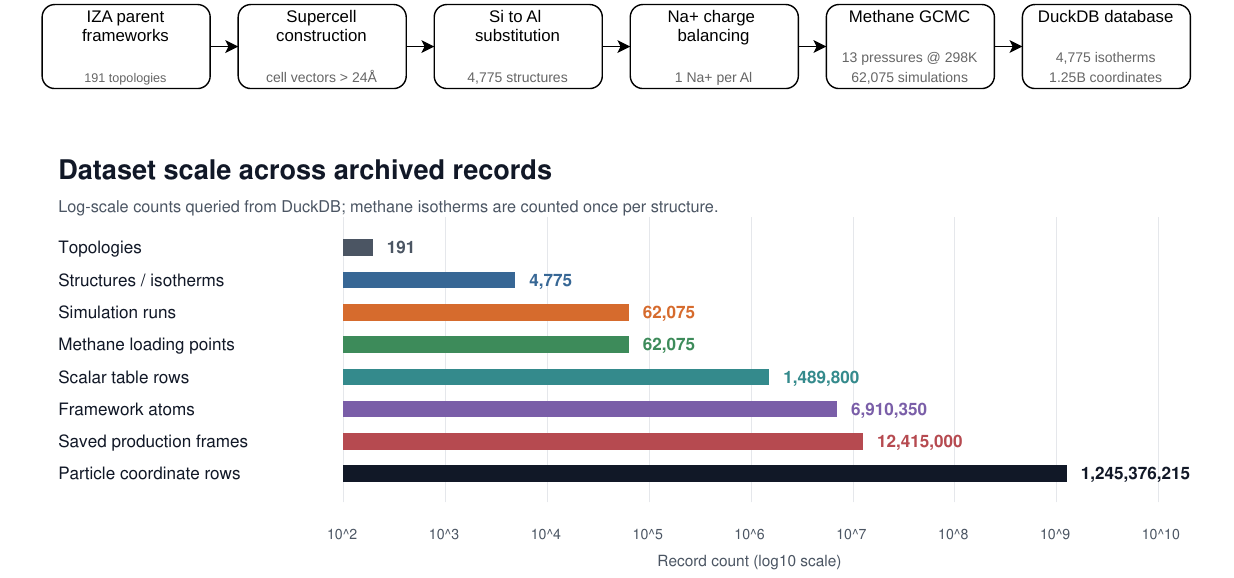}
\caption{Schematic overview of the dataset creation process and the main archived output scales. Parent zeolite frameworks were expanded into simulation cells, substituted with aluminium, charge-balanced with sodium, simulated with methane in the grand canonical ensemble, and stored in a processed DuckDB database containing scalar and coordinate-level records. The lower histogram shows the order-of-magnitude range of the archived tables, with methane isotherms counted once per framework realisation.}
\label{fig:workflow}
\end{figure}

Here, we present AdsZeo: a dataset of methane adsorption in aluminium-substituted zeolite frameworks. Configuration-level adsorption data can vary not only with pressure and loading, but also with the underlying pore topology, framework composition, adsorbate identity, and the presence of mobile extra-framework species. Capturing these sources of variation is important for machine learning models that aim to learn molecular distributions rather than only scalar adsorption properties. The present dataset provides a systematic collection across several of these factors. It contains 191 framework topologies and 25 aluminium-substituted framework realisations per topology, yielding 4,775 distinct framework realisations. Each framework realisation was simulated at 13 pressures between 0.1 and 100 bar, and the release stores scalar isotherm records together with production-frame coordinates of methane pseudo-atoms and sodium cations. The combination of diverse zeolite topologies, Al content and substitution patterns, charge-balancing sodium cations, and pressure-dependent methane configurations provides a substantial configuration-level resource for adsorption environments that are more structured than purely siliceous host--guest systems. An example of the distribution of methane and sodium cations in a saved adsorption configuration in a mordenite framework can be seen in Figure~\ref{fig:mor_dist}. The dataset is intended primarily for training and evaluating machine learning models for molecular configuration generation, density estimation, and simulation acceleration, while also providing a foundation for future extensions to mixtures, flexible adsorbates or frameworks, and other classes of porous materials. Figure~\ref{fig:workflow} summarises the dataset construction and the main archived data types.

\section{Methods}
The dataset was generated through a sequence of framework preparation, aluminium substitution, charge-balancing cation placement, GCMC simulation using RASPA \cite{dubbeldam2016raspa}, and post-processing steps. Parent all-silica framework structures representing the selected zeolite topologies were obtained from crystallographic data and converted into simulation-ready supercells. For each topology, multiple aluminium-substituted framework realisations were generated, with charge neutrality maintained by adding one extra-framework Na$^+$ cation per Al substitution. Methane adsorption was then simulated independently for each framework realisation on a fixed pressure grid. The processed dataset stores scalar adsorption records together with production-frame coordinates of methane and Na$^+$ cations.

\subsection{Framework preparation and aluminium substitution}

Parent all-silica framework structures were initialised from crystallographic data obtained from the International Zeolite Association database. \cite{baerlocher2007atlas}. Only framework topologies for which an accessible volume is reported in the IZA database were retained. This selection resulted in 191 parent framework topologies.

The crystallographic unit cells were processed with the \texttt{porran} package to generate simulation-ready atomic coordinates, unit-cell information, and periodic boundary conditions \cite{petkovic2025porran}. Each parent framework was expanded into a periodic supercell such that all lattice vector lengths exceeded 24~\AA. This criterion was chosen to satisfy the minimum-image requirement for the 12.0~\AA{} non-bonded cutoff used in the simulations. Framework atoms were treated as rigid during adsorption simulations, consistent with common high-throughput zeolite adsorption workflows. Framework coordinates are stored as fractional coordinates together with the corresponding unit-cell vectors.

Framework compositional disorder was introduced through explicit Si$\rightarrow$Al substitutions on tetrahedral sites. For each framework topology, 25 aluminium-substituted framework realisations were generated, spanning different Al substitution counts. The substitution level ranged from a single Al atom per supercell to a maximum aluminium fraction approaching 25\% of the available tetrahedral sites. Intermediate substitution counts were selected by linear interpolation between these bounds and rounded to the nearest integer. For each substitution count, Al sites were selected using a distance-maximisation procedure to obtain an approximately uniform spatial distribution of Al atoms throughout the simulation cell. Löwenstein's rule was not explicitly enforced, following the force-field treatment used for aluminosilicate zeolites in Ref.~\cite{romero2023adsorption}.

Each Si$\rightarrow$Al substitution introduces one negative framework charge. Charge neutrality was therefore maintained by adding one extra-framework Na$^+$ cation per Al atom. Initial Na$^+$ positions were generated randomly in the simulation cell. During the Monte Carlo simulations, Na$^+$ cations were treated as mobile particles, while their total number remained fixed by the framework charge balance. The resulting processed set contains 4,775 framework realisations from 191 zeolite topologies.

\subsection{Force-field model}

Methane adsorption was simulated using a classical, non-reactive force field assembled from previously validated zeolite adsorption models. Methane was represented with the united-atom hydrocarbon model of \citet{calero2004understanding}, in which each CH$_4$ molecule is described by a single neutral pseudo-atom. The aluminosilicate framework and Na$^+$ cation parameters were based on the zeolite force field of \citet{garcia2009transferable}, including the updated framework and cation parameters shown to describe adsorption in sodium-containing zeolites. For framework realisations containing Al--O--Al environments, the additional O$_\mathrm{aa}$ oxygen type introduced by \citet{romero2023adsorption} was included. This oxygen type was assigned the same Lennard-Jones interaction parameters as the other framework oxygen types, while its charge was adjusted according to the local aluminium environment.

Non-bonded interactions were described by Lennard-Jones 12-6 and Coulombic terms,

\begin{equation}
U_{ij}(r_{ij}) =
4\epsilon_{ij}
\left[
\left(\frac{\sigma_{ij}}{r_{ij}}\right)^{12}
-
\left(\frac{\sigma_{ij}}{r_{ij}}\right)^6
\right]
+
\frac{q_i q_j}{4\pi\varepsilon_0 r_{ij}} .
\end{equation}

Here, $r_{ij}$ is the distance between particles $i$ and $j$, $\epsilon_{ij}$ and $\sigma_{ij}$ are the Lennard-Jones energy and size parameters, and $q_i$ and $q_j$ are partial charges. Lennard-Jones interactions were shifted at a cutoff of 12.0~\AA{}, and tail corrections were not applied. Long-range electrostatic interactions were evaluated using Ewald summation. Lorentz--Berthelot mixing rules were used generally, with explicit pair interactions specified for the contributing adsorbate--framework and adsorbate--cation interactions. Framework atoms were treated as rigid, while methane pseudo-atoms and Na$^+$ cations were mobile during the Monte Carlo simulations.

Framework oxygen atoms were typed according to their local T--O--T environment. O denotes an oxygen atom connected to two Si atoms, O$_\mathrm{a}$ denotes an oxygen atom connected to one Si and one Al atom, and O$_\mathrm{aa}$ denotes an oxygen atom connected to two Al atoms. Silicon, aluminium, and oxygen framework atoms carried fixed partial charges. Silicon and aluminium did not contribute Lennard-Jones interactions directly in the simulations. The contributing Lennard-Jones pair interactions are listed in Table~\ref{tab:ff-pair-interactions}, and the pseudo-atom charges are listed in Table~\ref{tab:ff-charges}.

\begin{table}[htbp]
\centering
\caption{Contributing Lennard-Jones interactions used in the methane adsorption simulations. Interactions explicitly set to \texttt{none} in the simulation input files are not shown. Coulomb interactions are determined by the pseudo-atom charges in Table~\ref{tab:ff-charges} and evaluated using Ewald summation.}
\label{tab:ff-pair-interactions}
\small
\begin{tabular}{p{0.22\linewidth}p{0.22\linewidth}p{0.22\linewidth}p{0.18\linewidth}}
\toprule
Site $i$ & Site $j$ & $\epsilon_{ij}/k_\mathrm{B}$ / K & $\sigma_{ij}$ / \AA \\
\midrule
CH$_4$ & CH$_4$ & 158.5 & 3.72 \\
CH$_4$ & O / O$_\mathrm{a}$ / O$_\mathrm{aa}$  & 115.00 & 3.47 \\
CH$_4$ & Na$^+$ & 553.061 & 2.176 \\
Na$^+$ & O / O$_\mathrm{a}$ / O$_\mathrm{aa}$ & 23.0 & 3.4 \\
\bottomrule
\end{tabular}
\end{table}

\begin{table}[htbp]
\centering
\caption{Pseudo-atom charges used in the simulations. O, O$_\mathrm{a}$, and O$_\mathrm{aa}$ denote oxygen atoms connected to zero, one, or two Al atoms, respectively.}
\label{tab:ff-charges}
\small
\begin{tabular}{p{0.28\linewidth}p{0.42\linewidth}p{0.18\linewidth}}
\toprule
Site type & Description & Charge / $e$ \\
\midrule
Si & Framework silicon & 0.78598 \\
Al & Framework aluminium & 0.48598 \\
O & Si--O--Si framework oxygen & -0.39299 \\
O$_\mathrm{a}$ & Si--O--Al framework oxygen & -0.41384 \\
O$_\mathrm{aa}$ & Al--O--Al framework oxygen & -0.43469 \\
Na$^+$ & Extra-framework sodium cation & 0.38340 \\
CH$_4$ & United-atom methane pseudo-atom & 0.00000 \\
\bottomrule
\end{tabular}
\end{table}

\subsection{Grand canonical Monte Carlo simulations}

Methane adsorption was simulated with RASPA2 in the grand canonical ensemble \cite{dubbeldam2016raspa}. All simulations were performed at 298~K. Each framework realisation was simulated independently at 13 logarithmically spaced pressures between 0.1 and 100~bar. No continuation between pressures was used, and no repeated simulations with different random seeds were performed.

Each simulation consisted of 5,000 initialisation cycles followed by 200,000 production cycles. One Monte Carlo cycle was defined as $\max(20,N)$ trial moves, where $N$ is the instantaneous number of methane molecules and Na$^+$ cations in the simulation cell. Methane exchange with the gas reservoir was sampled through insertion and deletion moves. Configurational sampling used translation and reinsertion moves for methane and translation and reinsertion moves for Na$^+$ cations. The Na$^+$ population was fixed by framework charge balance, whereas the number of methane molecules fluctuated according to the imposed chemical potential. All move types were selected with equal probability.

During production, configurations were saved every 1,000 cycles, resulting in 200 saved production frames per simulation. Scalar adsorption quantities, including methane loading, molecule counts, sodium counts, and energy terms, were recorded from the same production runs. Simulations with zero methane loading were retained in the dataset because they are part of the equilibrium adsorption response at low pressure or in framework realisations with limited accessible pore volume.

\subsection{Coordinate extraction and data representation}

The released database stores production-frame coordinates of methane pseudo-atoms and Na$^+$ cations together with scalar adsorption records and framework metadata. Methane coordinates correspond to the centers of the united-atom methane model rather than to all-atom methane geometries. Adsorbate and cation coordinates are stored in Cartesian coordinates in \AA{} and are wrapped into the primary simulation cell. Framework coordinates are stored separately in fractional coordinates together with the unit-cell vectors, allowing the framework, methane, and Na$^+$ coordinates to be reconstructed in a common periodic cell.

Because the number of methane molecules fluctuates in the grand canonical ensemble, the number of methane pseudo-atoms varies between saved frames. The number of Na$^+$ cations is fixed for a given framework realisation and equals the number of Al substitutions. All records are linked through topology, structure, aluminium realisation, pressure, and frame identifiers. The processed data are stored in a DuckDB database to support efficient querying of scalar adsorption properties and configuration-level records.

\subsection{Simulation completeness and quality control}

All 4,775 framework realisations were simulated on the full 13-point pressure grid, yielding 62,075 adsorption simulations. All simulations completed successfully and are included in the released dataset. Post-processing checks verified the consistency of framework identifiers, pressure labels, methane and sodium counts, unit-cell metadata, and coordinate records. Charge neutrality was checked from the number of Al substitutions and Na$^+$ cations for each framework realisation.

Additional validation simulations were performed for selected framework-composition combinations for which experimental methane adsorption data were available. These comparisons were used to assess whether the force-field and simulation settings reproduce experimental adsorption behaviour for representative systems.

\section{Data Records}

The processed dataset has been deposited to Zenodo and is available from the AdsZeo Zenodo record \cite{adszeo}. The deposited processed-data file is the DuckDB database \texttt{zeolite\_stats.duckdb}, which contains eight tables: \texttt{structures}, \texttt{framework\_atoms}, \texttt{runs}, \texttt{cycle\_stats}, \texttt{positions}, \texttt{isotherms}, \texttt{batch\_ingest}, and \texttt{parse\_report}. The dataset should be cited using the Zenodo record.

\begin{table}[htbp]
\centering
\caption{Overview of the processed DuckDB tables. Record counts correspond to the deposited processed release.}
\label{tab:database-tables}
\begin{tabular}{p{0.22\linewidth}p{0.18\linewidth}p{0.50\linewidth}}
\toprule
Table & Record count & Contents \\
\midrule
\texttt{structures} & 4,775 & One row per framework realisation, including topology code, structure identifier, cell parameters, cell volume, atom counts, and Si/Al ratio. \\
\texttt{framework\_atoms} & 6,910,350 & Framework atom element labels, fractional coordinates, and charges. \\
\texttt{runs} & 62,075 & One row per pressure-specific simulation, including pressure, temperature, cycle settings, source paths, parse status, and final energy statistics. \\
\texttt{cycle\_stats} & 12,415,000 & Saved production-frame records with frame index, methane count, sodium count, total potential energy, energy components, and running adsorption quantities. \\
\texttt{positions} & 1,245,376,215 & Cartesian coordinates for methane pseudo-atoms and Na$^+$ cations by run, frame, species, and atom index. \\
\texttt{isotherms} & 1,489,800 & Absolute and excess methane loading records and reported uncertainties. \\
\texttt{batch\_ingest} & 51 & Batch-level ingestion metadata. \\
\texttt{parse\_report} & 0 & Parsing warnings, errors, or skipped files. \\
\bottomrule
\end{tabular}
\end{table}

The \texttt{structures} table contains one row per framework realisation. It stores the zeolite topology code, structure identifier, unit-cell parameters, supercell information, cell volume, atom counts, and Si/Al ratio. The \texttt{framework\_atoms} table stores the host-framework atoms associated with each framework realisation, including element labels, fractional coordinates, and charges. The \texttt{runs} table stores one row per pressure-specific simulation and includes the thermodynamic state point, simulation settings, source paths, parse status, and final reported energy statistics.

The \texttt{cycle\_stats} table contains 200 saved records per simulation run. These records include frame index, methane and sodium counts, total potential energy, energy components, and running adsorption quantities in multiple units. The \texttt{positions} table stores Cartesian coordinates for methane pseudo-atoms and Na$^+$ cations by run, frame, species, and atom index. The \texttt{isotherms} table stores absolute and excess methane loading values and reported uncertainties. The \texttt{batch\_ingest} and \texttt{parse\_report} tables document ingestion status and parsing issues.

\begin{figure}[htbp]
\includegraphics[width=\textwidth]{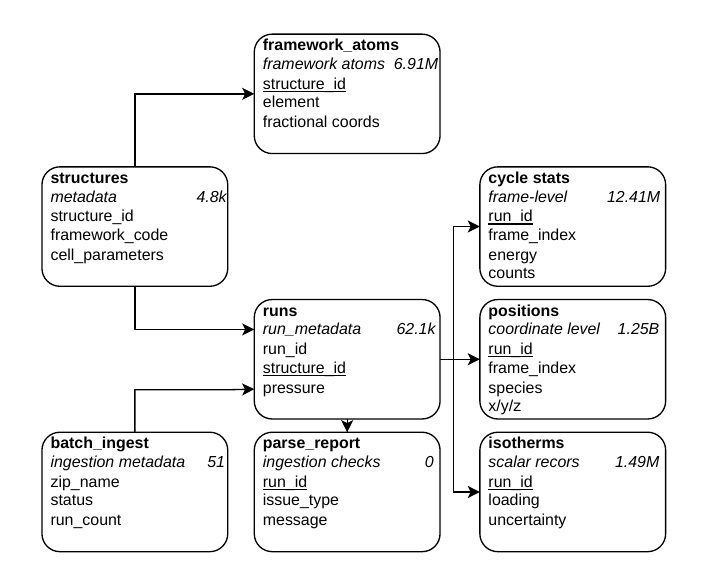}
\caption{Relational structure of the processed database. Structure identifiers link framework metadata and framework atoms to pressure-specific simulation runs, while run identifiers link each simulation to saved cycle statistics, scalar isotherm records, and methane and sodium coordinate records.}
\label{fig:data-model}
\end{figure}

Figure~\ref{fig:data-model} illustrates the relationships between the database tables and the identifiers required to join framework metadata, simulation runs, scalar records, and coordinate records.

\section{Technical Validation}

The dataset was validated for production-sampling stability, consistency between saved coordinate records and scalar adsorption quantities, geometric validity of the saved configurations, and pressure-dependent adsorption behaviour. These checks were designed to verify that the saved production-frame coordinates are consistent with equilibrated GCMC sampling and with the scalar adsorption records generated by RASPA. The processed database was also checked for completeness and internal consistency. All 62,075 simulations, corresponding to 4,775 framework realisations at 13 pressures, were successfully parsed. Each simulation contains the expected 200 saved production records, giving 12,415,000 saved production frames in total. A comparison with experimental methane adsorption data is provided as an additional external validation of the simulated adsorption behaviour.

\subsection{Monte Carlo production stability}

The stability of the saved production sampling was assessed using block-wise statistics and representative production traces. For each simulation, the 200 saved production records were divided into four consecutive blocks of 50 records. The average methane count in each block was compared with the average over all 200 saved production records. This provides a direct check for residual drift within the saved production segment and for simulations in which the production average would be dominated by a non-stationary part of the production sequence. The same block-wise analysis was also applied to the total potential energy traces as a qualitative check of production stability.

Across the full dataset, the signed methane-count block deviations were centred at zero, with a 5 - 95\% interval from $-0.745$ to 0.750 CH$_4$ per unit cell. The absolute block deviation had a mean of 0.337 CH$_4$ per unit cell, a median of 0.250 CH$_4$ per unit cell, and a 95th percentile of 0.960 CH$_4$ per unit cell. The largest observed absolute block deviation was 7.935 CH$_4$ per unit cell. Relative block deviations were evaluated only for simulations with an average methane count of at least 4 CH$_4$ per unit cell, so that near-zero-loading simulations were not dominated by large relative errors coming from a deviation of 1 methane molecule per unit cell. For this subset, the absolute relative block deviation had a mean of 1.46\%, a median of 0.79\%, and a 95th percentile of 5.26\%.

\begin{figure}[htbp]
\centering
\includegraphics[width=\textwidth]{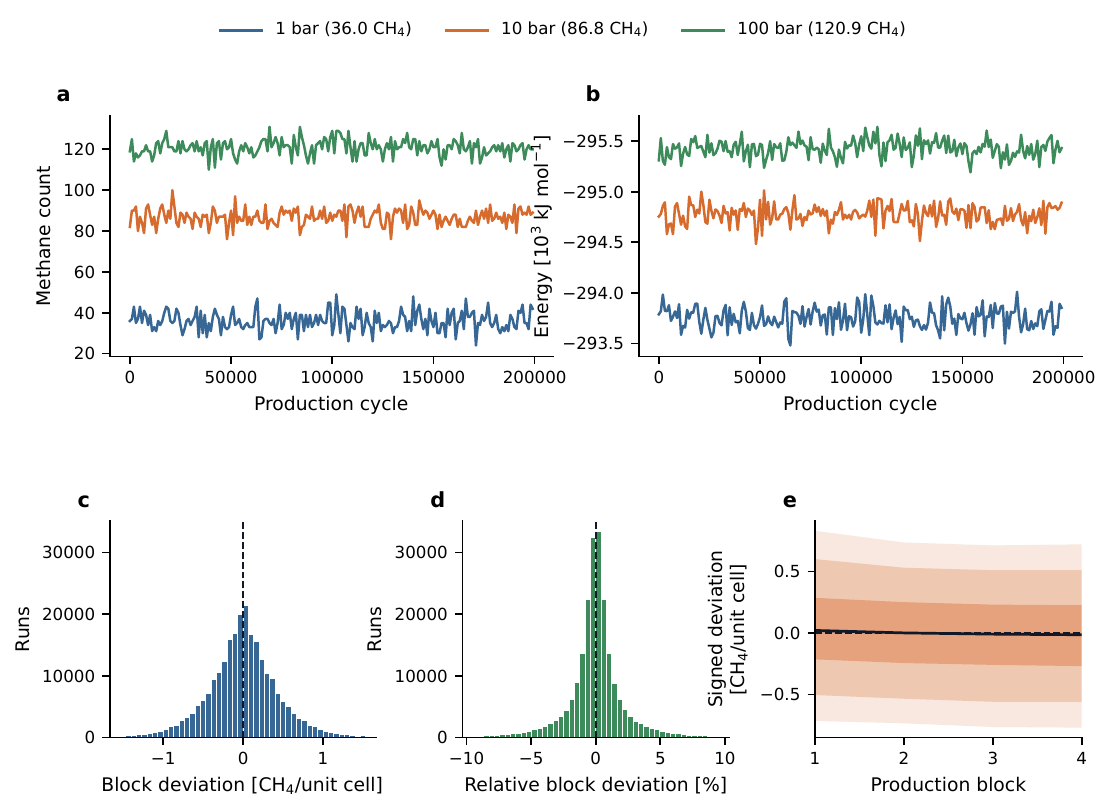}
\caption{Monte Carlo production stability. 
(a) Representative methane-count traces for simulations at 1, 10, and 100 bar; mean methane counts over the saved production records are given in parentheses. 
(b) Corresponding total potential-energy traces. 
(c) Distribution of signed block-average methane-count deviations relative to the full saved-production average. 
(d) Distribution of absolute relative block deviations for simulations with an average methane count of at least 4 CH$_4$ per unit cell. 
(e) Signed block deviations for the four consecutive production blocks, showing that the deviations remain centred close to zero throughout the saved production segment.}
\label{fig:technical-validation-convergence}
\end{figure}

Figure~\ref{fig:technical-validation-convergence} shows representative production traces and the full-dataset block statistics. The example traces at 1, 10, and 100 bar fluctuate around stable mean methane counts of 32.3, 50.5, and 60.0 CH$_4$, respectively, without visible long-time drift over the saved production records. The corresponding potential-energy traces show stationary fluctuations around pressure-dependent mean energies. The block-resolved signed deviations remain centred close to zero for all four production blocks, indicating that the saved production records do not show a appreciable, systematic first-to-last-block drift across the dataset.

\subsection{Consistency between saved configurations and scalar adsorption records}

A central purpose of the dataset is to provide coordinate-level adsorption samples together with scalar adsorption quantities. We therefore checked whether the scalar methane loading can be reconstructed from the saved production-frame coordinates. For each simulation, methane pseudo-atoms were counted in each of the 200 saved frames and averaged over the saved production records. The resulting coordinate-derived loading was then compared with the scalar loading reported by RASPA.

This comparison validates that the saved methane coordinates, frame identifiers, pressure identifiers, and scalar adsorption records refer to the same production simulations. Exact equality is not expected, because the coordinate-derived loading is calculated from the 200 saved production frames, whereas the RASPA scalar average is accumulated internally over the full production run of 200,000 cycles. The comparison therefore measures consistency between the saved coordinate samples and the scalar production averages, rather than identity between two independently stored quantities.

\begin{figure}[htbp]
\centering
\includegraphics[width=\textwidth]{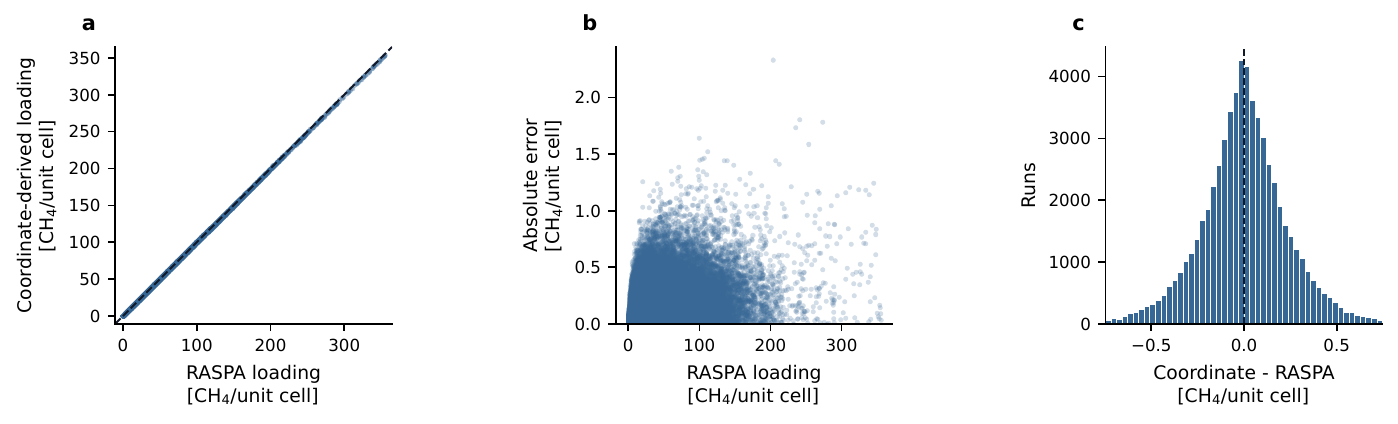}
\caption{Consistency between saved coordinate records and scalar adsorption records. 
(a) Parity plot comparing methane loading reconstructed from saved production-frame coordinates with the RASPA-reported methane loading. The dashed line indicates exact agreement. 
(b) Absolute loading error as a function of RASPA-reported loading. 
(c) Distribution of signed loading differences, defined as coordinate-derived loading minus RASPA-reported loading.}
\label{fig:coordinate-loading-consistency}
\end{figure}

The coordinate-derived loadings agreed closely with the RASPA-reported loadings over the full loading range of the dataset (Figure~\ref{fig:coordinate-loading-consistency}). Across all 62,075 simulations, the mean absolute deviation was 0.1794 CH$_4$ per unit cell, the root-mean-square deviation was 0.2439 CH$_4$ per unit cell, and the maximum absolute deviation was 2.3317 CH$_4$ per unit cell. The signed error distribution was centred near zero, indicating no appreciable systematic offset between the coordinate-derived and RASPA-reported loadings.

\subsection{Geometric validity of saved configurations}

The saved production-frame coordinates were validated to ensure that the processed configurations are physically meaningful under periodic boundary conditions. During post-processing, methane pseudo-atoms and Na$^+$ cations were wrapped into the primary simulation cell. The number of methane coordinate records in each saved frame was checked against the corresponding saved methane count, and the number of Na$^+$ cations was checked against the number of Al substitutions in the associated framework realisation. Frames with zero methane loading were also verified to contain no methane coordinates while retaining the expected Na$^+$ cation records. No non-finite coordinates were found in the processed dataset.

\begin{figure}[htbp]
\centering
\includegraphics[width=\textwidth]{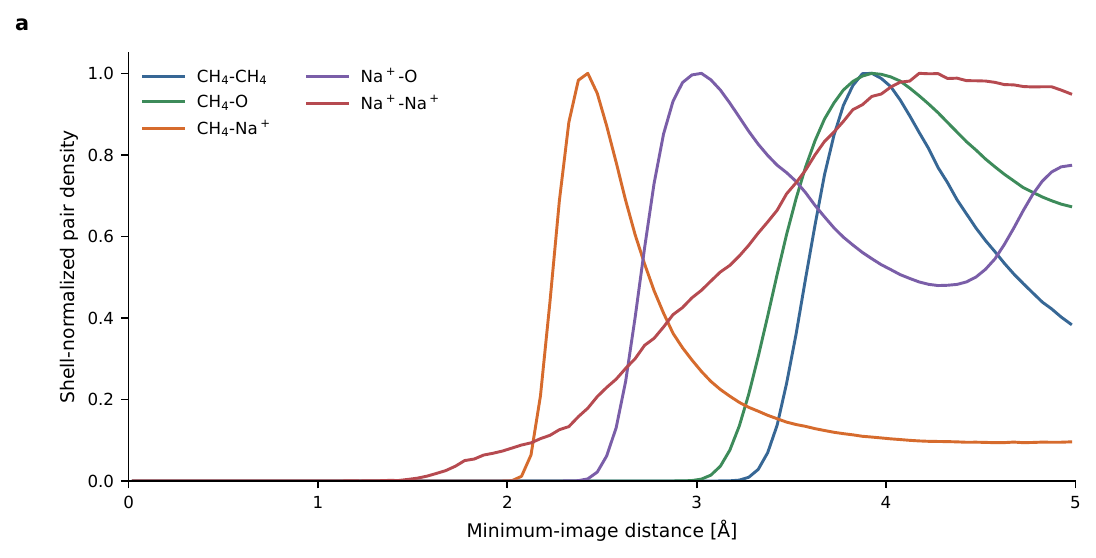}
\caption{Geometric validation of saved configurations. Shell-normalised minimum-image pair-distance distributions were computed for the 100~bar simulations using every 25th saved production frame. The checked pair types were CH$_4$--CH$_4$, CH$_4$--framework O, CH$_4$--Na$^+$, Na$^+$--framework O, and Na$^+$--Na$^+$. Each distribution is normalised by its maximum value to compare the onset and characteristic distance ranges of the different pair types. The absence of density at very short distances provides a check against unphysical overlaps, duplicated coordinates, and coordinate-wrapping artefacts.}
\label{fig:minimum-distance-validation}
\end{figure}

As an additional check for coordinate artefacts, minimum-image pair distances were computed for representative high-loading configurations. This analysis was performed for the 100~bar simulations using every 25th saved production frame, since these configurations provide the most stringent test for short-range overlaps involving adsorbed methane. The checked pair types were CH$_4$--CH$_4$, CH$_4$--framework O, CH$_4$--Na$^+$, Na$^+$--framework O, and Na$^+$--Na$^+$. The resulting shell-normalised pair-distance distributions are shown in Figure~\ref{fig:minimum-distance-validation}. The distributions show no intensity at unphysically short distances, indicating that the checked configurations do not show evidence of duplicated atoms, failed coordinate wrapping, or malformed coordinate records. The shortest pronounced feature is observed for CH$_4$--Na$^+$ separations, followed by Na$^+$--O and methane-containing framework distances at larger separations. The broad Na$^+$--Na$^+$ distribution reflects the range of cation arrangements present across the generated framework realisations.

\subsection{Pressure-dependent adsorption behaviour}

The scalar adsorption records were further checked for physically consistent pressure-dependent behaviour. For each framework realisation, the methane loading was evaluated across the full pressure grid and the resulting isotherm was tested for monotonicity. All simulated isotherms increased monotonically with pressure, as expected for equilibrium methane adsorption in the pressure range considered here. This check provides a simple but important validation that the pressure labels, loading records, and parsed simulation outputs are internally consistent.

\begin{figure}[htbp]
\centering
\includegraphics[width=\textwidth]{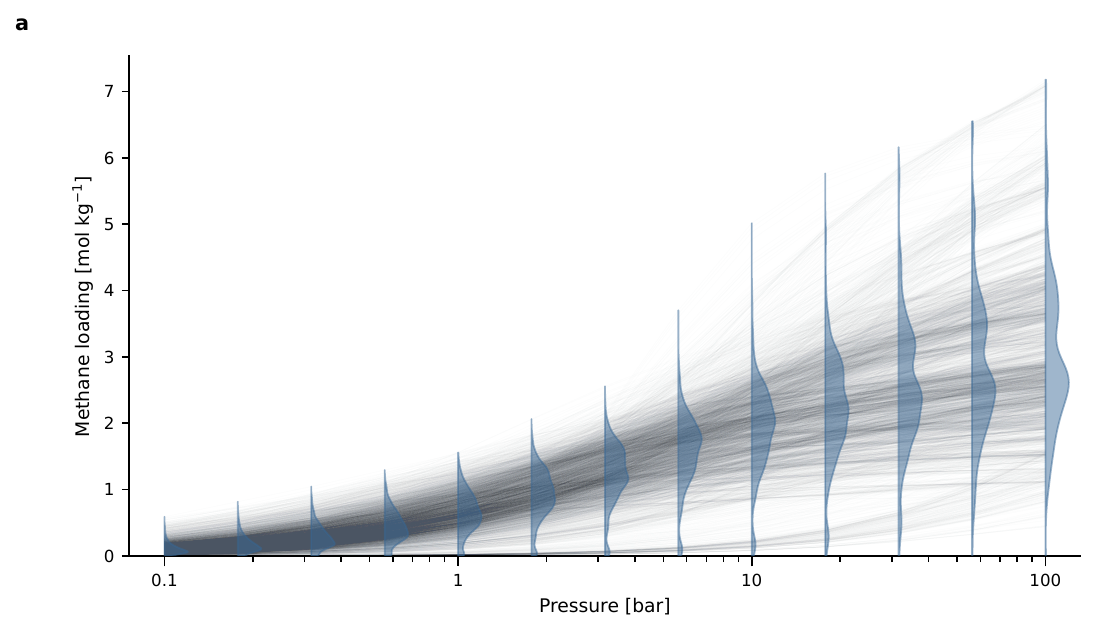}
\caption{Pressure-dependent methane loading distribution across the dataset. Thin grey lines show individual simulated isotherms, while the pressure-resolved density distributions summarise the spread of methane loading values at each pressure point. The loading range broadens with increasing pressure, reflecting variation in framework topology, Al distribution, Na$^+$ cation arrangement, and accessible adsorption volume. All isotherms were verified to increase monotonically with pressure.}
\label{fig:isotherm-distribution}
\end{figure}

Figure~\ref{fig:isotherm-distribution} summarises the distribution of methane loadings across the dataset as a function of pressure. The individual isotherms show a broad range of adsorption responses, reflecting differences in framework topology, Al distribution, Na$^+$ cation arrangement, and available adsorption volume. At low pressure, most framework realisations exhibit near-zere methan loading, while the spread in loading increases with pressure as framework-specific adsorption capacities and adsorption-site distributions become more important. The pressure-resolved loading distributions therefore confirm that the dataset contains a diverse set of adsorption behaviours while remaining free of obvious unphysical trends such as negative loadings or decreasing isotherms.

\subsection{Comparison with experimental methane adsorption data}

As an external validation of the simulated adsorption behaviour, methane isotherms were compared with experimental data for three sodium-exchanged zeolite systems: Na-ZSM-5, NaY, and NaX \cite{dunne1996calorimetric,talu1993effect,chkhaidze1985adsorption}. For each system, ten framework realisations were generated from the corresponding parent pure-silica IZA framework by introducing the appropriate number of Al substitutions and charge-balancing Na$^+$ cations. The number of Al atoms per unit cell used for the comparison were 3 Al atoms for Na-ZSM-5, 60 Al atoms for NaY, and 86 Al atoms for NaX. Al substitutions were generated subject to the non-Loewenstein constraint, and methane adsorption isotherms were simulated using the same force-field and GCMC settings as used for the dataset-generation runs.

\begin{figure}[htbp]
\centering
\includegraphics[width=\textwidth]{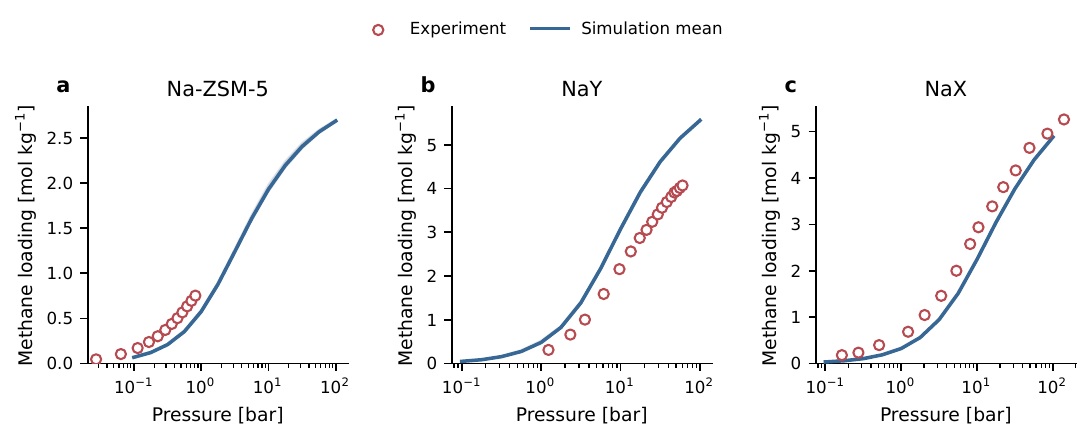}
\caption{Comparison between simulated and experimental methane adsorption isotherms for selected sodium-exchanged zeolites. Experimental data are shown as open red circles and the mean simulated loading over ten generated framework realisations is shown as a blue line. Comparisons are shown for (a) Na-ZSM-5, (b) NaY, and (c) NaX. The simulations use the same force-field and GCMC settings as the dataset-generation runs. Deviations from experiment are expected because the simulated structures were generated from ideal pure-silica frameworks with substituted Al and Na$^+$ cations, rather than from sample-specific experimentally refined structures.}
\label{fig:experimental-validation}
\end{figure}

The comparison is shown in Figure~\ref{fig:experimental-validation}. The simulations reproduce the overall pressure-dependent uptake trends and give adsorption magnitudes that are broadly consistent with experiment. Agreement is particularly good for NaX over the available pressure range, where the simulated and experimental curves follow a similar loading increase with pressure. For Na-ZSM-5, the experimental data cover only the low-pressure region relative to the simulated pressure range; within this overlap, the simulations capture the low loading regime and then predict the continued increase toward higher-pressure saturation. For NaY, the simulated curve follows the experimental trend but gives higher loadings at elevated pressure.

Quantitative agreement with experiment is not expected to be exact in this comparison, because the simulated structures were generated by substituting Al into ideal pure-silica frameworks rather than by using experimentally refined or fully optimised structures for each specific zeolite sample. Differences in unit-cell dimensions, Al siting, cation positions, sample composition, and experimental conditions can all affect methane uptake. The comparison should therefore be interpreted as an external consistency check rather than a direct force-field refit. Overall, the agreement supports that the dataset captures realistic methane adsorption behaviour while retaining the systematic, generated-structure design needed for coordinate-resolved machine-learning applications.

\section{Usage Notes}

The processed database can be queried directly with DuckDB from Python, R, or the DuckDB command-line client. A companion notebook, \path{working_with_adszeo_data.ipynb}, provides executable examples for inspecting the schema, constructing a run-level index, selecting a leakage-aware machine-learning split, retrieving one framework realisation and one saved production frame, converting framework fractional coordinates to Cartesian coordinates, and packaging a frame as arrays for downstream models. The notebook is intended as the recommended starting point for users who want to build machine-learning datasets from the deposited DuckDB file.

Typical workflows should begin by selecting \texttt{structure\_id} and \texttt{run\_id} values from \texttt{structures} and \texttt{runs}. These identifiers can then be joined to \texttt{isotherms} for scalar adsorption quantities, to \texttt{cycle\_stats} for saved production-frame statistics, and to \texttt{positions} for methane and Na$^+$ coordinates. Because \texttt{positions} is the largest table, coordinate queries should be restricted to a manifest of selected \texttt{run\_id}, \texttt{frame\_index}, pressure, and species values rather than scanning the full table. Users who only need scalar adsorption properties can work with \texttt{structures}, \texttt{runs}, \texttt{cycle\_stats}, and \texttt{isotherms} without loading coordinate records.

For machine-learning applications, data splits should be defined before retrieving coordinates. Random frame-level splits are usually inappropriate because frames from the same production run are correlated, pressure points from the same framework realisation share the same host framework, and framework realisations derived from the same topology are related. Conservative benchmarks for configuration generation or density estimation should split by \texttt{framework\_code} to test transfer to unseen topologies. Less stringent benchmarks may split by \texttt{structure\_id} to test interpolation across aluminium-substitution realisations, but frames from the same \texttt{run\_id} should remain in the same split.

The coordinate representation should be chosen to match the modelling task. Framework atoms are stored in fractional coordinates together with the unit-cell parameters, while methane pseudo-atoms and Na$^+$ cations are stored as wrapped Cartesian coordinates in \AA{}. Periodic boundary conditions should therefore be handled explicitly when computing distances, voxel grids, density fields, or equivariant graph inputs. The number of methane pseudo-atoms varies between frames in the grand canonical ensemble, whereas the number of Na$^+$ cations is fixed for a given framework realisation by the number of aluminium substitutions. Zero-loading frames are valid samples and should be retained unless the modelling task explicitly excludes them.

The scalar isotherm records and the saved coordinate frames describe the same production simulations but are not identical averages. Scalar adsorption quantities reported by RASPA are accumulated over the full production run, while coordinate-derived quantities are estimated from the 200 saved frames. For generative modelling, density estimation, or simulation-acceleration tasks, pressure, temperature, framework topology, aluminium content, cell parameters, and Na$^+$ coordinates should be treated as conditioning information. The saved configurations are Monte Carlo samples from rigid-framework simulations with the force field described above and should not be interpreted as experimental trajectories.

\section{Data Availability}

The dataset is available from Zenodo at \url{https://doi.org/10.5281/zenodo.21445386} \cite{adszeo}.

\section{Code Availability}

All code used to submit the RASPA simulations, parse the RASPA output, construct the processed DuckDB database, analyse the dataset, and generate the manuscript figures is available at \url{https://github.com/Rachna-R/AdsZeo}. The repository includes the extraction scripts, figure-generation scripts, and the RASPA simulation template used for the production calculations.
The Zenodo repository at \url{https://zenodo.org/records/21445386}
contains a companion usage notebook at \path{working_with_adszeo_data.ipynb} showing how to work with the dataset.

RASPA input preparation was performed using Python 3.13.1, pymatgen 2025.5.2, and NumPy 2.3.5. Python 3.13.1 and QCG-PilotJob 0.14.2 were used to submit the simulation jobs to the computing cluster, and the adsorption simulations were performed using RASPA 2.0.47. Data extraction, analysis, and figure generation were performed using Python 3.14.4, DuckDB 1.5.4, pandas 3.0.3, NumPy 2.4.6, and Matplotlib 3.11.0. These packages and software can be found at the following URLs:

\begin{itemize}
\item Python: \url{https://www.python.org/}
\item RASPA: \url{https://iraspa.org/raspa}
\item QCG-PilotJob: \url{https://qcg-pilotjob.readthedocs.io/en/develop/}
\item pymatgen: \url{https://pymatgen.org/}
\item NumPy: \url{https://numpy.org/}
\item matplotlib: \url{https://matplotlib.org/}
\item duckdb: \url{https://duckdb.org/}
\item pandas: \url{https://pandas.pydata.org/}
\end{itemize}

\section*{Funding}
This work used the Dutch national e-infrastructure with the support of the
SURF Cooperative using grant no. EINF-18440. This work was funded by the European Union’s Horizon Europe as part of the SUPERVAL project (The Sustainable Photo-Electrochemical Valorization of flue gases), grant agreement no. 101115456.

\end{document}